\documentclass[a4paper,11pt]{article}
\usepackage[usenames]{color}

\usepackage{amsfonts,amssymb}
\usepackage{theorem}
\usepackage{cancel}
\usepackage[dvips]{graphicx}
\newcommand{\su}[2]{\stackrel{#1}{#2}}

\def\g5{\gamma_{_\chi}}

\newcommand{\comment}[1]{}

\begin{document}
\comment
{
  \rightline{\Huge gamma\_5\_paper\_II\_v6.tex}
\par
\rightline{Pisa, February 12, 2015}
\par
\rightline{revised Pisa, March 25,  2015 at 17.00}
}
\vskip 1.0 truecm
\Large
\bf
\begin{center}{Managing $\gamma_5$ in Dimensional Regularization II:\\
the Trace with more $\gamma_5$'s
\footnote{\tt This work is supported in part by funds provided by the U.S. Department
of Energy (D.O.E.) under cooperative research agreement \#DE FG02-05ER41360} }
\end{center}
\par
\normalsize
\rm

\rm 
\large
\vskip 1.1 truecm
\centerline{
Ruggero~Ferrari \footnote{e-mail: {\tt ruggferr@mit.edu}}}
\normalsize
\medskip
\begin{center}
Center for Theoretical Physics\\
Laboratory for Nuclear Science\\
and Department of Physics\\
Massachusetts Institute of Technology\\
Cambridge, Massachusetts 02139\\
\comment{
and\\ INFN, Sezione di Milano\\
via Celoria 16, I-20133 Milano, Italy\\}
(MIT-CTP4643 March 2015)
\end{center}
%
%

\normalsize
\bf
\centerline{Abstract}

\rm
\begin{quotation}
In the present paper we evaluate the anomaly
for the abelian axial current in a non abelian chiral gauge theory,
by using dimensional regularization. This amount to
formulate a procedure for managing  traces  
with more than one $\gamma_5$.
\par
The suggested procedure obeys Lorentz covariance 
and cyclicity, at variance with previous approaches
(e.g. the celebrated 't Hooft and Veltman's where Lorentz
is violated)
\par
The result of the present paper is a further step forward
in the program initiated by a previous work on the
traces involving a single $\gamma_5$. The final goal is 
an unconstrained definition of $\gamma_5$ in dimensional
regularization. Here, in the evaluation of the anomaly,
we profit of the axial current conservation  equation,  when
radiative corrections are neglected. This kind of tool
is not always exploited in field theories with $\gamma_5$, 
e.g. in the use of dimensional regularization of infrared
and collinear divergences..

\end{quotation}
PACS: 11.10.Gh, 
11.30.Rd, 
11.40.Ha 
\newpage
\section{Introduction}
\label{sec:intr}
In paper I (\cite{Ferrari:2014jqa}) we 
solved the problem of defining  
the trace of gamma's
with zero or one $\gamma_5$ in generic $D$ dimensions
\cite{'tHooft:1972fi}-\cite{Cicuta:1972jf}, 
by using an integral representation. The $\gamma_5$ problem has
been widely discussed in the literature 
\cite{Rosenberg:1962pp}-\cite{Tsai:2009hp}.
\par
The new representation sets the
rules for managing the algebra in a Lorentz covariant
formalism, consistent with the cyclicity of the trace.
The ABJ anomaly \cite{Adler:1969gk}-\cite{Bardeen:1969md} 
and the LFE (Local Functional Equation) 
\cite{Ferrari:2005ii}-\cite{Bettinelli:2007kc}
associated to the abelian local chiral transformations
have been verified by explicit calculations. 
\par
In the present paper we consider the case of a trace
with more than one $\gamma_5$, that frequently occurs in actual 
Feynman amplitude calculations.
There is a further cogent reason to consider such a
case, i.e. the need to formulate local chiral non abelian gauge
transformations, as in the electroweak model. Were
it not possible to do it in a consistent way, then 
the $\gamma_5$ manipulation in generic dimension
would be of limited significance.
\par
In this work we go through the explicit calculation
of the divergence of the abelian axial current
\begin{eqnarray}
\partial_\mu J_\mu^5
\label{intr.0}
\end{eqnarray}
up to one loop correction in a $SU(2)$ nonabelian 
chiral (massless) theory. We use dimensional 
regularization and the limit $D=4$ is taken.
\par
We make some assumptions, hoping that they are mutually
consistent:
\par\noindent
1. Gamma's and $\gamma_\chi$ (our $\gamma_5$
in generic $D$) form an associative algebra. 
\par
We study
the generic trace where the Lorentz indices are all
contracted with vectors (e.g. momenta and polarization vectors)
and tensors (as $\delta_{\mu\nu}$)
\begin{eqnarray}
Tr(p)\equiv Tr\Big(\dots\gamma_\chi\dots\gamma_{\mu_j}\dots\gamma_\chi\dots
\gamma_{\mu_k}\dots\gamma_{\mu_{k'}}\dots\Big)
\dots p_{\mu_j}\dots \delta_{\mu_k\mu_{k'}}\dots\, .
\label{intr.01}
\end{eqnarray}
Then our {\sl Ansatz} is that:
\par\noindent
2.  In a neighborhood of $D=4$ the trace
admits an expansion
\begin{eqnarray}
Tr(p)
= \sum _{h=0} A_h(p) (D-4)^h,
\label{intr.02}
\end{eqnarray}
where $A_h(p)$ are Lorentz invariants in $D=4$ dimensions (
the tensor $\varepsilon_{\mu\nu\rho\sigma}$ might be present).
\par\noindent
3.
The limit $D=4$ is smooth. For instance
\begin{eqnarray}
\{\gamma_\chi,\gamma_\mu\} = {\cal O}(D-4),~ \forall \mu.
\label{intr.03}
\end{eqnarray}
To our opinion the integral representation of the trace
with  zero or one $\gamma_\chi$, thoroughly  studied
in I, can be extended to the case of multiple $\gamma_\chi$.
However we have not been able yet to continue our integral
representation for any number of $\gamma_\chi$ to non integer $D$; 
i.e. the manipulations, requiring an integer 
$D$, provide little help in order to extend the results to non integer $D$.
For these reasons and for sake of brevity and conciseness
we do not discuss here the extension to multiple
$\gamma_\chi$ of the results in I. Instead we
manipulate in a formal way the gamma's, assuming that
they exist somehow.
\par
For instance the trace $Tr\big(\gamma_\chi\gamma_\alpha\gamma_\beta
\gamma_\mu\gamma_\nu\gamma_\rho\gamma_\sigma\big)$ need not to be
given. In the evaluation of the anomaly only the following quantity
is required
\begin{eqnarray}&&
Tr\big(\big\{\gamma_\chi,~\gamma_\alpha\big\}\gamma_\beta
\gamma_\mu\gamma_\nu\gamma_\rho\gamma_\sigma\big)
\nonumber\\&&
=Tr\big(\gamma_\chi\big\{\gamma_\alpha,~\gamma_\beta
\gamma_\mu\gamma_\nu\gamma_\rho\gamma_\sigma\big\}\big).
\label{intr.03.1}
\end{eqnarray}
\par
The strategy for evaluating the trace with many 
$\gamma_\chi$ turns out to be very simple at the one-loop level.
\begin{enumerate}
\item
We move around, inside a trace,  a $\gamma_\chi$ by introducing 
the anticommutator. For instance 
\begin{eqnarray}&&
\gamma_\chi\gamma_\mu = -\gamma_\mu\gamma_\chi
+\{\gamma_\chi,\gamma_\mu\}
\nonumber\\&&
[\gamma_\chi,\gamma_\chi] =0.
\label{intr.1-1}
\end{eqnarray}
\item Once the anticommutator $\{\gamma_\chi,\gamma_\mu\}$ 
is introduced into the trace
we get only ${\cal O}(D-4)$ quantities or of higher order
in $D-4$.
\item If we need only terms of first order in  $D-4$  and 
$\{\gamma_\chi,\gamma_\mu\}$ is present,
 then we can use the $D=4$ algebra in the subsequent
manipulation (e.g. $\gamma_\chi^2=1$ and $\{\gamma_\chi,\gamma_\mu\}=0$).
\item Eventually the trace contains at most one $\gamma_\chi$,
if $\{\gamma_\chi,\gamma_\mu\}$ is present and if only first $D-4$ order
terms are required.
\end{enumerate}
 Trace with at most one $\gamma_\chi$ have been dealt in I.
\par
To summarize, the method is very simple and straightforward.
Once the ${\cal O}(D-4)$ factor is introduced into the
trace via a single anticommutator $\{\gamma_\chi,\gamma_\mu\}$,
the $D=4$ na\"{\i}ve
algebra  can be used
\begin{eqnarray}&&
\gamma_\chi \not\!p_1\dots \not\! p_k \gamma_\chi
\to (-)^k \not\!p_1\dots \not\! p_k 
\nonumber\\&&
\gamma_\chi^2 = 1\, .
\label{intr.1}
\end{eqnarray}
However powers of $\gamma_\chi$ need some care as it is
discussed in Section \ref{sec:alg}.
\par
In the present paper we apply the above outlined method 
to the evaluation of the anomaly present in
the operator (\ref{intr.0}). First we organize
all contributions to the operator $\partial_\mu J_\mu^5$
in such a way that they identically vanish if one uses
the na\"{\i}ve $D=4$ algebra (i.e.
if poles in $D=4$  are neglected).
 With this procedure we can factorize
$\{\gamma_\chi,\gamma_\mu\}$ in the trace. 
Then the evaluation of the anomaly is straightforward.
%
\section{More Algebraic Properties}
\label{sec:alg}
The algebra of $\gamma_\chi$ with the other gamma's
is not know. Thus the algebraic manipulations go
around this difficulty. As an example, used frequently
in I, we quote the following identity
\begin{eqnarray}&&
Tr \Bigl(\{\gamma_\alpha,\gamma_\chi\}\gamma_\rho
\gamma_\beta\gamma_\sigma\gamma_\iota\gamma_\mu \Bigr) \delta_{\alpha\iota}
=
Tr \Bigl(\gamma_\chi\{\gamma_\alpha,\gamma_\rho
\gamma_\beta\gamma_\sigma\gamma_\iota\gamma_\mu\} \Bigr) \delta_{\alpha\iota}
\nonumber\\&&
=(2-D)Tr \Bigl(\gamma_\chi\gamma_\rho
\gamma_\beta\gamma_\sigma\gamma_\mu \Bigr) 
+
Tr \Bigl(\gamma_\chi
\nonumber\\&&
\Big[(6 -D)\gamma_\rho\gamma_\beta\gamma_\sigma
-4 (\delta_{\rho\beta}\gamma_\sigma-\delta_{\rho\sigma}\gamma_\beta+\delta_{\sigma\beta}\gamma_\rho)  \Big]\gamma_\mu
 \Bigr)
\nonumber\\&&
=Tr \Bigl(\gamma_\chi
\Big[2(4 -D)\gamma_\rho\gamma_\beta\gamma_\sigma
-4 (\delta_{\rho\beta}\gamma_\sigma-\delta_{\rho\sigma}\gamma_\beta+\delta_{\sigma\beta}\gamma_\rho)  \Big]\gamma_\mu
 \Bigr)
\label{pro.2}
\end{eqnarray}
which is zero both for $D=4$ and $D=2$, as it should be.
\par
Here we list some rules and some {\sl caveat}.
It should be reminded that the na\"{\i}ve $D=4$
algebra  can be used only under the protection
of a ${\cal O}(D-4)$ factor in the trace.
For instance 
\begin{eqnarray}
\gamma_\chi^2 = 1
\label{alg.1}
\end{eqnarray}
cannot be used under all circumstances. Here is an
example of some unpleasant difficulty 
%
\begin{eqnarray}&&
Tr ~\Bigl(\gamma_\mu \su{}{\gamma}_\chi
\gamma_\alpha  \gamma_\rho \su{}{\gamma}_\chi\gamma_\beta
\gamma_\sigma
\gamma_\iota\su{}{\gamma}_\chi
\Bigr)
\nonumber\\&&
= - Tr ~\Bigl(\su{}{\gamma}_\chi\gamma_\mu 
\gamma_\alpha  \gamma_\rho \su{}{\gamma}_\chi\gamma_\beta
\gamma_\sigma
\gamma_\iota\su{}{\gamma}_\chi
\Bigr)
+ Tr ~\Bigl(\{\gamma_\mu,~\su{}{\gamma}_\chi \}
\gamma_\alpha  \gamma_\rho \su{}{\gamma}_\chi\gamma_\beta 
\gamma_\sigma
\gamma_\iota\su{}{\gamma}_\chi
\Bigr)
\nonumber\\&&
= - Tr ~\Bigl(\su{}{\gamma}_\chi\gamma_\mu 
\gamma_\alpha  \gamma_\rho \su{}{\gamma}_\chi\gamma_\beta
\gamma_\sigma
\gamma_\iota\su{}{\gamma}_\chi
\Bigr)
- Tr ~\Bigl(\{\gamma_\mu,~ \su{}{\gamma}_\chi\}
\gamma_\alpha  \gamma_\rho \gamma_\beta 
\gamma_\sigma
\gamma_\iota
\Bigr)
\nonumber\\&&
= - Tr ~\Bigl(\su{}{\gamma}_\chi\su{}{\gamma}_\chi\gamma_\mu
\gamma_\alpha  \gamma_\rho \su{}{\gamma}_\chi\gamma_\beta
\gamma_\sigma
\gamma_\iota
\Bigr)
- Tr ~\Bigl(\{\gamma_\mu,~ \su{}{\gamma}_\chi\}
\gamma_\alpha  \gamma_\rho \gamma_\beta 
\gamma_\sigma
\gamma_\iota
\Bigr)
\label{foot.1}
\end{eqnarray}
But also
\begin{eqnarray}&&
Tr~ \Bigl(\gamma_\mu \su{}{\gamma}_\chi
\gamma_\alpha  \gamma_\rho \su{}{\gamma}_\chi\gamma_\beta
\gamma_\sigma
\gamma_\iota\su{}{\gamma}_\chi
\Bigr)
=  Tr ~\Bigl(\su{}{\gamma}_\chi\gamma_\mu \su{}{\gamma}_\chi
\gamma_\alpha  \gamma_\rho \su{}{\gamma}_\chi\gamma_\beta
\gamma_\sigma
\gamma_\iota
\Bigr)
\nonumber\\&&
= - Tr ~\Bigl(\gamma_\mu \su{}{\gamma}_\chi\su{}{\gamma}_\chi
\gamma_\alpha  \gamma_\rho \su{}{\gamma}_\chi\gamma_\beta
\gamma_\sigma
\gamma_\iota
\Bigr)
+ Tr ~\Bigl(\{\su{}{\gamma}_\chi,~\gamma_\mu\} \su{}{\gamma}_\chi
\gamma_\alpha  \gamma_\rho \su{}{\gamma}_\chi\gamma_\beta
\gamma_\sigma
\gamma_\iota
\Bigr)
\nonumber\\&&
= - Tr ~\Bigl(\gamma_\mu \su{}{\gamma}_\chi\su{}{\gamma}_\chi
\gamma_\alpha  \gamma_\rho \su{}{\gamma}_\chi\gamma_\beta 
\gamma_\sigma
\gamma_\iota
\Bigr)
+ Tr ~\Bigl(\{\su{}{\gamma}_\chi,~\gamma_\mu\}
\gamma_\alpha \gamma_\rho \gamma_\beta\gamma_\sigma\gamma_\iota\Bigr)
\nonumber\\&&
\label{foot.2}
\end{eqnarray}
Thus eqs. (\ref{foot.1}) and  (\ref{foot.2}) are in contradiction
is we use $\gamma_\chi^2=1$. The last identity can be used only
inside a trace where a ${\cal O}(D-4)$ term already is present.
\par
Moreover one can easily derive
\begin{eqnarray}&&
\nonumber\\&&
 Tr ~\Bigl(\big[\gamma_\mu, ~\su{}{\gamma}_\chi
 \su{}{\gamma}_\chi\big]
\gamma_\alpha  \gamma_\rho \su{}{\gamma}_\chi\gamma_\beta\gamma_\sigma
\gamma_\iota\Bigr)
=2 Tr ~\Bigl(\{\su{}{\gamma}_\chi,~\gamma_\mu\}
\gamma_\alpha  \gamma_\rho \gamma_\beta
\gamma_\sigma\gamma_\iota\Bigr)
\nonumber\\&&
\label{foot.3}
\end{eqnarray}
which shows once more how $\gamma_\chi^2$ is difficult object
to deal with.
\par
In some cases we can use $\gamma_\chi^2=1$ in proximity
of $D=4$. In our calculation we encounter two
cases of this sort.
\begin{eqnarray}&&
 Tr\Big(\Big[\gamma_\mu,\gamma_\chi^2\Big]
~\gamma_\alpha\gamma_\rho
\gamma_\beta~
\Big)
\nonumber\\&&
 Tr\Big(\Big[\gamma_\mu,\gamma_\chi^2\Big]
~\gamma_\alpha\gamma_\rho
\gamma_\beta~ \gamma_\sigma~\gamma_\iota
\Big)
\label{alg.3}
\end{eqnarray}
We can easily prove that around $D=4$ they can be neglected.
For instance
\begin{eqnarray}&&
 Tr\Big(\Big[\gamma_\mu,\gamma_\chi^2\Big]
\gamma_\alpha\gamma_\rho
\gamma_\beta~
\Big)\delta_{\mu\alpha}
=(D-4) Tr\Big(\gamma_\chi^2\gamma_\rho
\gamma_\beta~\Big)+ 4 \delta_{\rho\beta} Tr\Big(\gamma_\chi^2\Big)
\nonumber\\&&
- DTr\Big(\gamma_\chi^2\gamma_\rho
\gamma_\beta~\Big)=0
\label{alg.4}
\end{eqnarray}
and 
\begin{eqnarray}&&
 Tr\Big(\Big[\gamma_\mu,\gamma_\chi^2\Big]
\gamma_\alpha\gamma_\rho
\gamma_\beta\gamma_\sigma\gamma_\iota
\Big)\delta_{\mu\alpha}
=(D-8) Tr\Big(\gamma_\chi^2\gamma_\rho
\gamma_\beta~\gamma_\sigma~\gamma_\iota\Big)
\nonumber\\&&
+ 4 \delta_{\rho\beta} Tr\Big(\gamma_\chi^2\gamma_\sigma\gamma_\iota \Big)
- 4\delta_{\rho\sigma} Tr\Big(\gamma_\chi^2\gamma_\beta\gamma_\iota \Big)
+4\delta_{\rho\iota} Tr\Big(\gamma_\chi^2\gamma_\beta\gamma_\sigma \Big)
\nonumber\\&&
+ 4 \delta_{\beta\sigma} Tr\Big(\gamma_\chi^2\gamma_\rho\gamma_\iota \Big)
-4 \delta_{\beta\iota} Tr\Big(\gamma_\chi^2\gamma_\rho\gamma_\sigma \Big)
+4 \delta_{\sigma\iota} Tr\Big(\gamma_\chi^2\gamma_\rho\gamma_\beta \Big)
\nonumber\\&&
- DTr\Big(\gamma_\chi^2\gamma_\rho
\gamma_\beta\gamma_\sigma\gamma_\iota\Big)=0
\label{alg.5}
\end{eqnarray}
are compatible with both traces in eq. (\ref{alg.3})
being zero at $D\sim 4$.
\section{The Anomaly of isoscalar $J^5_\mu$ 
in Chiral Nonabelian Gauge Theories}
\label{sec:anona}
In chiral theory every vertex carries a factor 
\begin{eqnarray}
\frac{1}{2}(1+\gamma_\chi).
\label{anona.1}
\end{eqnarray}
The triangular graph gives the amplitude
(a factor $i^3$ from fermion propagators, a factor $i^2$ 
for  interaction vertexes and a factor $-1$ from fermion
loop. Total $-i$)
\begin{eqnarray}&&
T_{\mu\rho\sigma}(k,p)={-\frac{i}{4}}
\int \frac{d^D q}{(2\pi)^D}~
\frac{1}
{
(q-k)^2 q^2 (q+p)^2
}
\nonumber\\&&
 Tr ~\Bigl(\gamma_\mu~\gamma_\chi
(q-k)_\alpha\gamma_\alpha  ~\gamma_\rho(1+\gamma_\chi) ~q_\beta\gamma_\beta~ 
\gamma_\sigma~(1+\gamma_\chi)
(q+p)_\iota\gamma_\iota
\Bigr).
\label{anona.2}
\end{eqnarray}
The crossed graph will be added later on.
The trace on the internal group indices contributes by a factor
\begin{eqnarray}
Tr(t_a t_b)= \frac{1}{2} \delta_{ab}.
\label{anona.4}
\end{eqnarray}
Eventually we consider the divergence of the current
(eq. (\ref{intr.0}))
\begin{eqnarray}&&
i~
(p+k)_\mu  T_{\mu\rho\sigma}(k,p)
={\frac{1}{4}}
(q+p-q+k)_\mu
\nonumber\\&&
\int \frac{d^D q}{(2\pi)^D}~
\frac{
 Tr ~\Bigl(\gamma_\mu~\gamma_\chi 
(q-k)_\alpha\gamma_\alpha  ~\gamma_\rho(1+\gamma_\chi) ~q_\beta\gamma_\beta~ 
\gamma_\sigma~(1+\gamma_\chi)
(q+p)_\iota\gamma_\iota
\Bigr)
}
{
(q-k)^2 q^2 (q+p)^2
}
\nonumber\\&&
={\frac{1}{4}}\int \frac{d^D q}{(2\pi)^D}~
\Bigg\{-\frac{  Tr ~\Bigl((q-k)_\mu\gamma_\mu \gamma_\chi(q-k)_\alpha
\gamma_\alpha}
{
(q-k)^2}
\nonumber\\&&
\frac{
 ~\gamma_\rho(1+\gamma_\chi) ~q_\beta\gamma_\beta~ 
\gamma_\sigma~(1+\gamma_\chi)
(q+p)_\iota\gamma_\iota
\Bigr)
}{ q^2 (q+p)^2}
\nonumber\\&&
+\frac{
 Tr ~\Bigl(\gamma_\chi
(q-k)_\alpha\gamma_\alpha  ~\gamma_\rho(1+\gamma_\chi) ~q_\beta\gamma_\beta~ 
\gamma_\sigma~(1+\gamma_\chi)
\Bigr)
}
{(q-k)^2 q^2 }
\Bigg\}
\label{anona.2.1}
\end{eqnarray}
The crossed graph yields
\begin{eqnarray}&&
i~
(p+k)_\mu  T_{\mu\sigma\rho}(p,k)
={\frac{1}{4}}
(q+k-q+p)_\mu
\nonumber\\&&
\int \frac{d^D q}{(2\pi)^D}~
\frac{
 Tr ~\Bigl(\gamma_\mu~\gamma_\chi 
(q-p)_\alpha\gamma_\alpha  ~\gamma_\sigma(1+\gamma_\chi) ~q_\beta\gamma_\beta~ 
\gamma_\rho~(1+\gamma_\chi)
(q+k)_\iota\gamma_\iota
\Bigr)
}
{
(q-p)^2 q^2 (q+k)^2
}
\nonumber\\&&
={\frac{1}{4}}\int \frac{d^D q}{(2\pi)^D}~
\Bigg\{-\frac{  Tr ~\Bigl((q-p)_\mu\gamma_\mu \gamma_\chi(q-p)_\alpha
\gamma_\alpha}
{
(q-p)^2}
\nonumber\\&&
\frac{
 ~\gamma_\sigma(1+\gamma_\chi) ~q_\beta\gamma_\beta~ 
\gamma_\rho~(1+\gamma_\chi)
(q+k)_\iota\gamma_\iota
\Bigr)
}{ q^2 (q+k)^2}
\nonumber\\&&
+\frac{
 Tr ~\Bigl(\gamma_\chi
(q-p)_\alpha\gamma_\alpha  ~\gamma_\sigma(1+\gamma_\chi) ~q_\beta\gamma_\beta~ 
\gamma_\rho~(1+\gamma_\chi)
\Bigr)
}
{(q-p)^2 q^2 }
\Bigg\}.
\label{anona.2.2}
\end{eqnarray}
We shift the variable  $q\to q - k$ in the {first} integral
and  $q\to q + p$ in the {second} of eq. (\ref{anona.2.2}).
\begin{eqnarray}&&
i~
(p+k)_\mu  T_{\mu\sigma\rho}(p,k)
\nonumber\\&&
={\frac{1}{4}}\int \frac{d^D q}{(2\pi)^D}~
\Bigg\{-\frac{  Tr ~\Bigl((q-k-p)_\mu\gamma_\mu \gamma_\chi(q-k-p)_\iota
\gamma_\iota}
{
(q-k-p)^2}
\nonumber\\&&
\frac{
 ~\gamma_\sigma(1+\gamma_\chi) ~(q-k)_\alpha\gamma_\alpha~ 
\gamma_\rho~(1+\gamma_\chi)
q_\beta\gamma_\beta
\Bigr)
}{ q^2 (q-k)^2}
\nonumber\\&&
+\frac{
 Tr ~\Bigl(\gamma_\chi
q_\beta\gamma_\beta  ~\gamma_\sigma(1+\gamma_\chi) ~(q+p)_\iota\gamma_\iota~ 
\gamma_\rho~(1+\gamma_\chi)
\Bigr)
}
{(q+p)^2 q^2 }
\Bigg\}.
\label{anona.2.3}
\end{eqnarray}
By inspection one sees that the first term in eq. (\ref{anona.2.1})
cancels the second term in eq. (\ref{anona.2.3}) if one uses 
na{\"\i}vely the algebra in $D=4$. The same happens 
to the second term in eq. (\ref{anona.2.1})
with the first term in eq. (\ref{anona.2.3}). Our strategy
is to find the anomaly in the lack of these cancellations,
when radiative corrections are taken into account.
As an example we deal with one of these two cases.
Thus we have
\begin{eqnarray}&&
{\frac{1}{4}}\int \frac{d^D q}{(2\pi)^D}~
\Bigg\{-\frac{  Tr ~\Bigl((q-k)_\mu\gamma_\mu \gamma_\chi(q-k)_\alpha
\gamma_\alpha}
{
(q-k)^2}
\nonumber\\&&
\frac{
 ~\gamma_\rho(1+\gamma_\chi) ~q_\beta\gamma_\beta~ 
\gamma_\sigma~(1+\gamma_\chi)
(q+p)_\iota\gamma_\iota
\Bigr)
}{ q^2 (q+p)^2}
\nonumber\\&&
+\frac{
 Tr ~\Bigl(\gamma_\chi
q_\beta\gamma_\beta  ~\gamma_\sigma(1+\gamma_\chi) ~(q+p)_\iota\gamma_\iota~ 
\gamma_\rho~(1+\gamma_\chi)
\Bigr)
}
{(q+p)^2 q^2 }
\Bigg\}
\nonumber\\&&
=
{\frac{1}{4}}\int \frac{d^D q}{(2\pi)^D}~
\Bigg\{-\frac{  Tr ~\Bigl((q-k)_\mu\Big(-\gamma_\chi\gamma_\mu+\{\gamma_\mu, \gamma_\chi\}
\Big)(q-k)_\alpha
\gamma_\alpha}
{
(q-k)^2}
\nonumber\\&&
\frac{
 ~\gamma_\rho(1+\gamma_\chi) ~q_\beta\gamma_\beta~ 
\gamma_\sigma~(1+\gamma_\chi)
(q+p)_\iota\gamma_\iota
\Bigr)
}{ q^2 (q+p)^2}
\nonumber\\&&
+\frac{
 Tr ~\Bigl(
q_\beta\gamma_\beta  ~\gamma_\sigma(1+\gamma_\chi) ~(q+p)_\iota\gamma_\iota~ 
\Big(-\gamma_\chi\gamma_\rho~+\{\gamma_\chi,\gamma_\rho\}\Big)(1+\gamma_\chi)
\Bigr)
}
{(q+p)^2 q^2 }
\Bigg\}
\nonumber\\&&
=
{\frac{1}{4}}\int \frac{d^D q}{(2\pi)^D}~
\Bigg\{- Tr ~\Bigl(\{\gamma_\mu, \gamma_\chi\}\gamma_\alpha
\gamma_\rho(1+\gamma_\chi)\gamma_\beta~ 
\gamma_\sigma~(1+\gamma_\chi)\Bigr)
\nonumber\\&&
\frac{ (q-k)_\mu(q-k)_\alpha
q_\beta (q+p)_\iota\gamma_\iota
}{ q^2 (q+p)^2(q-k)^2}
\nonumber\\&&
+
 Tr ~\Bigl(
\gamma_\beta  ~\gamma_\sigma(1+\gamma_\chi)\gamma_\iota~ 
\{\gamma_\chi,\gamma_\rho\}(1+\gamma_\chi)
\Bigr)
\frac{q_\beta ~(q+p)_\iota
}
{(q+p)^2 q^2 }
\Bigg\}
\label{anona.2.1+3}
\end{eqnarray}
Eq. (\ref{anona.2.1+3}) gives a contribution
to the triangular graph anomaly. The cross term will be 
be added later on. Noticeable is the emerging inside
the trace of the factors
$\{\gamma_\mu, \gamma_\chi\}$ and $\{\gamma_\chi,\gamma_\rho\}$
of order ${\cal O}(D-4)$.

\subsection{Reduction of $\gamma_\chi$'s}
\label{sec:red}
We proceed to remove all $\gamma_\chi$'s in eq. (\ref{anona.2.1+3})
where it is possible. The guiding idea is that the presence in the trace 
of the factors
$\{\gamma_\mu, \gamma_\chi\}$, which is of order ${\cal O}(D-4)$,
allows us the use 
\begin{eqnarray}&&
\{ \gamma_\chi, ~\gamma_\nu\}=0, ~
\forall \nu
\nonumber\\&&
\gamma_\chi^2 = 1
\label{anona.2.4-1}
\end{eqnarray}
for all the other remaining $\gamma_\chi$'s.
The generic
value $D$ is kept throughout the computation and the limit
$\gamma_\chi \to \gamma_5 $ is taken as a last step of the algebraic
manipulation of  $\{\gamma_\mu, \gamma_\chi\}$.
\par
Thus we consider the gamma content of
the first term in eq. (\ref{anona.2.1})
\begin{eqnarray}&&
Tr ~\Bigl(\{\gamma_\mu,~\gamma_\chi\} 
\gamma_\alpha  \gamma_\rho(1+\gamma_\chi) \gamma_\beta
\gamma_\sigma(1+\gamma_\chi)
\gamma_\iota
\Bigr)
\nonumber\\&&
=Tr ~\Bigl(\{\gamma_\mu,~\gamma_\chi\}  
\gamma_\alpha  \gamma_\rho \gamma_\beta 
\gamma_\sigma
\gamma_\iota
\Bigr)
\nonumber\\&&
+Tr ~\Bigl(\{\gamma_\mu,~\gamma_\chi\} 
\gamma_\alpha  \gamma_\rho \gamma_\beta 
\gamma_\sigma\gamma_\chi
\gamma_\iota
\Bigr)
\nonumber\\&&
+Tr ~\Bigl(\{\gamma_\mu,~\gamma_\chi\} 
\gamma_\alpha  \gamma_\rho\gamma_\chi \gamma_\beta 
\gamma_\sigma
\gamma_\iota
\Bigr)
\nonumber\\&&
+Tr ~\Bigl(\{\gamma_\mu,~\gamma_\chi\}
\gamma_\alpha  \gamma_\rho \gamma_\chi \gamma_\beta 
\gamma_\sigma\gamma_\chi
\gamma_\iota
\Bigr)
\label{anona.2.4}
\end{eqnarray}
The first term in the RHS of eq. (\ref{anona.2.4})
gives
\begin{eqnarray}
Tr ~\Bigl((\{\gamma_\mu,~\gamma_\chi\}
\gamma_\alpha  \gamma_\rho \gamma_\beta 
\gamma_\sigma
\gamma_\iota
\Bigr)
=Tr ~\Bigl(\gamma_\chi\{ \gamma_\mu,
\gamma_\alpha  \gamma_\rho \gamma_\beta 
\gamma_\sigma
\gamma_\iota\}
\Bigr)
\label{anona.2.5}
\end{eqnarray}
The fourth term in the RHS of eq. (\ref{anona.2.4})
gives 
\begin{eqnarray}&&
Tr ~\Bigl( \{\gamma_\mu,~\su{}{\gamma}_\chi\}
\gamma_\alpha  \gamma_\rho\su{}{\gamma}_\chi  \gamma_\beta 
\gamma_\sigma \su{}{\gamma}_\chi
\gamma_\iota
\Bigr)
\nonumber\\&&
=
Tr ~\Bigl( \{\gamma_\mu,~\su{}{\gamma}_\chi\}
\gamma_\alpha  \gamma_\rho \gamma_\beta 
\gamma_\sigma \gamma_\iota
\Bigr)
=
Tr ~\Bigl(\su{}{\gamma}_\chi \{\gamma_\mu,~
\gamma_\alpha  \gamma_\rho \gamma_\beta 
\gamma_\sigma \gamma_\iota\}
\Bigr)
\label{anona.2.6}
\end{eqnarray}
Finally the first and the fourth together yield
\begin{eqnarray}&&
Tr ~\Bigl(\{\gamma_\mu,~\gamma_\chi \}
\gamma_\alpha  \gamma_\rho \gamma_\beta 
\gamma_\sigma
\gamma_\iota
\Bigr)   +  
Tr ~\Bigl( \{\gamma_\mu,~\su{}{\gamma}_\chi\}
\gamma_\alpha  \gamma_\rho\su{}{\gamma}_\chi  \gamma_\beta 
\gamma_\sigma\su{}{\gamma_\chi }
\gamma_\iota
\Bigr)
\nonumber\\&&
=2Tr ~\Bigl(\gamma_\chi \{\gamma_\mu,~
\gamma_\alpha  \gamma_\rho \gamma_\beta 
\gamma_\sigma \gamma_\iota\}
\Bigr).
\label{anona.2.7}
\end{eqnarray}
\par
Now we consider the second and third terms in eq.
(\ref{anona.2.4}) i.e. where an even number of $\gamma_\chi$ is present.
\begin{eqnarray}&&
Tr ~\Bigl(\{\gamma_\mu,~\su{}{\gamma}_\chi\}
\gamma_\alpha  \gamma_\rho \gamma_\beta 
\gamma_\sigma\su{}{\gamma}_\chi
\gamma_\iota
\Bigr)
+Tr ~\Bigl( \{\gamma_\mu,~\su{}{\gamma}_\chi\}
\gamma_\alpha  \gamma_\rho \su{}{\gamma}_\chi\gamma_\beta 
\gamma_\sigma
\gamma_\iota
\Bigr)
\nonumber\\&&
=
Tr ~\Bigl( [\gamma_\mu,~\gamma^2_\chi]
\gamma_\alpha  \gamma_\rho \gamma_\beta 
\gamma_\sigma
\gamma_\iota
\Bigr)=0
\label{anona.2.4.1}
\end{eqnarray}
according to the arguments of Section \ref{sec:alg}.
\par
The same  analysis has to be performed on the gamma
content of the second term in eq. (\ref{anona.2.1+3})
which should match the first eq. (\ref{anona.2.1+3}) or 
of eq. (\ref{anona.2.1})
\begin{eqnarray}&&
Tr ~\Bigl(\{\gamma_\chi,~\gamma_\rho\} (1+\gamma_\chi) \gamma_\beta\gamma_\sigma~ 
(1+\gamma_\chi) \gamma_\iota 
\Bigr)  
\nonumber\\&&
=
Tr ~\Bigl(\{\gamma_\chi,~ \gamma_\rho\} \gamma_\beta\gamma_\sigma~
\gamma_\iota 
\Bigr)  
\nonumber\\&&
+Tr ~\Bigl(\{\gamma_\chi, ~\gamma_\rho\}\gamma_\chi  \gamma_\beta\gamma_\sigma~ 
\gamma_\iota 
\Bigr)  
\nonumber\\&&
+Tr ~\Bigl(\{\gamma_\chi,~ \gamma_\rho\}  \gamma_\beta\gamma_\sigma~ 
\gamma_\chi\gamma_\iota 
\Bigr)  
\nonumber\\&&
+Tr ~\Bigl(\{\gamma_\chi, ~\gamma_\rho\}\gamma_\chi  \gamma_\beta\gamma_\sigma~ 
\gamma_\chi\gamma_\iota 
\Bigr)  
\label{anona.2.8}
\end{eqnarray}
We elaborate on the single terms as for eq. (\ref{anona.2.4})
\begin{eqnarray}&&
=
Tr ~\Bigl(\gamma_\chi\{ \gamma_\rho, \gamma_\beta\gamma_\sigma~
\gamma_\iota \}
\Bigr)  
\nonumber\\&&
+Tr ~\Bigl([\gamma_\rho,\gamma^2_\chi ]  \gamma_\beta\gamma_\sigma~ 
\gamma_\iota 
\Bigr)  
\nonumber\\&&
+Tr ~\Bigl(\{\gamma_\chi, \gamma_\rho\}\gamma^2_\chi  \gamma_\beta\gamma_\sigma~ 
\gamma_\iota 
\Bigr)  
\label{anona.2.9}
\end{eqnarray}
where now all terms are zero for $D\sim 4$.
\par
Finally the only surviving of the gamma's algebra is the term in the
RHS of eq. (\ref{anona.2.7})
\begin{eqnarray}&&
i~
(p+k)_\mu ( T_{\mu\rho\sigma}(k,p)+T_{\mu\sigma\rho}(p,k))
\nonumber\\&&
={\frac{1}{4}}\int \frac{d^D q}{(2\pi)^D}~
\Bigg\{
\frac{2Tr ~\Bigl(\gamma_\chi\{ \gamma_\mu,
\gamma_\alpha  ~\gamma_\rho ~\gamma_\beta~ 
\gamma_\sigma~
\gamma_\iota\}
\Bigr)(q-k)_\mu(q-k)_\alpha q_\beta(q+p)_\iota
}{ (q-k)^2q^2 (q+p)^2}
\nonumber\\&&
+(k\leftrightarrow p)(\rho\leftrightarrow\sigma )\Bigg\}.
\label{anona.2.10.0}
\end{eqnarray}
%

\subsection{Symmetric Integration}
\label{sec:sym}
We use Feynman parameterization in order to perform
a symmetric integration over $q$
\begin{eqnarray}&&
i~
(p+k)_\mu ( T_{\mu\rho\sigma}(k,p)+T_{\mu\sigma\rho}(p,k))
\nonumber\\&&
=
{\frac{1}{4}}  2 \int_0^1 dx\int_0^x dy \int \frac{d^D q}{(2\pi)^D}~
\Bigg\{
2Tr ~\Bigl(\gamma_\chi\{ \gamma_\mu,
\gamma_\alpha  \gamma_\rho \gamma_\beta 
\gamma_\sigma
\gamma_\iota\}
\Bigr)
\nonumber\\&&
\frac{
(q+r-k)_\mu(q+r-k)_\alpha (q+r)_\beta(q+r+p)_\iota
}{ (q^2-\Delta)^3}
\nonumber\\&&
+(k\leftrightarrow p)(\rho\leftrightarrow\sigma )\Bigg\}
\label{anona.2.11}
\end{eqnarray}
with
\begin{eqnarray}
r_\nu \equiv (yk-xp+yp)_\nu.
\label{anona.2.12}
\end{eqnarray}
We keep only those terms that survive in the limit $D=4$
\begin{eqnarray}&&
i~
(p+k)_\mu ( T_{\mu\rho\sigma}(k,p)+T_{\mu\sigma\rho}(p,k))
\nonumber\\&&
=
{\frac{1}{4D}}  2 \int_0^1 dx\int_0^x dy \int \frac{d^D q}{(2\pi)^D}~
\frac{q^2
}{ (q^2-\Delta)^3}
\nonumber\\&&
\Bigg\{
4Tr\gamma_\chi\Big[ ~\Bigl(\big(\delta_{\mu\alpha} \gamma_\rho \gamma_\beta
\gamma_\sigma
\gamma_\iota
-\delta_{\mu\rho } \gamma_\alpha\gamma_\beta 
\gamma_\sigma
\gamma_\iota
+\delta_{\mu \beta} \gamma_\alpha\gamma_\rho 
\gamma_\sigma
\gamma_\iota
\nonumber\\&&
-\delta_{\mu\sigma } \gamma_\alpha\gamma_\rho 
\gamma_\beta
\gamma_\iota
+\delta_{\mu\iota } \gamma_\alpha\gamma_\rho 
\gamma_\beta
\gamma_\sigma 
\Bigr)\Big]
\nonumber\\&&
\Big ( \delta_{\mu\alpha}r_\beta(r+p)_\iota 
+ \delta_{\mu\beta}(r-k)_\alpha(r+p)_\iota
+\delta_{\mu\iota}(r-k)_\alpha r_\beta
\Big)
\nonumber\\&&
+(k\leftrightarrow p)(\rho\leftrightarrow\sigma )\Bigg\}
\nonumber\\&&
=
{\frac{1}{4D}}  2 \int_0^1 dx\int_0^x dy \int \frac{d^D q}{(2\pi)^D}~
\frac{q^2
}{ (q^2-\Delta)^3}
\nonumber\\&&
\Bigg\{
4Tr\gamma_\chi \Big[\gamma_\rho \gamma_\beta 
\gamma_\sigma\gamma_\iota\Bigl(Dr_\beta p_\iota +(r-k)_\beta(r+p)_\iota
+k_\beta r_\iota\Bigr)
\nonumber\\&&
- \gamma_\rho \gamma_\beta 
\gamma_\sigma\gamma_\iota\Bigl(r_\beta p_\iota -(r-k)_\beta(r+p)_\iota
-k_\beta r_\iota\Bigr)
\nonumber\\&&
+ \gamma_\rho \gamma_\beta 
\gamma_\sigma\gamma_\iota\Bigl(-r_\beta p_\iota- D(r-k)_\beta(r+p)_\iota
+k_\beta r_\iota \Bigr)
\nonumber\\&&
- \gamma_\rho \gamma_\beta 
\gamma_\sigma\gamma_\iota\Bigl(r_\beta p_\iota- (r-k)_\beta(r+p)_\iota
-k_\beta r_\iota \Bigr)
\nonumber\\&&
+ \gamma_\rho \gamma_\beta 
\gamma_\sigma\gamma_\iota\Bigl(-r_\beta p_\iota +(r-k)_\beta(r+p)_\iota
-D k_\beta r_\iota \Bigr)
\Big]
+(k\leftrightarrow p)(\rho\leftrightarrow\sigma )\Bigg\}
\nonumber\\&&
=
  \frac{2}{{D}} \int_0^1 dx\int_0^x dy \int \frac{d^D q}{(2\pi)^D}~
\frac{q^2
}{ (q^2-\Delta)^3}(D-4)
\nonumber\\&&
Tr\Big(\gamma_\chi \gamma_\rho \gamma_\beta 
\gamma_\sigma\gamma_\iota\Big)\Bigl(r_\beta p_\iota {-}(r-k)_\beta(r+p)_\iota
-k_\beta r_\iota\Bigr)
\nonumber\\&&
+(k\leftrightarrow p)(\rho\leftrightarrow\sigma )
\nonumber\\&&
=
   \frac{2}{{D}} \int_0^1 dx\int_0^x dy \int \frac{d^D q}{(2\pi)^D}~
\frac{q^2
}{ (q^2-\Delta)^3}(D-4)
\nonumber\\&&
Tr\Big(\gamma_\chi \gamma_\rho \gamma_\beta 
\gamma_\sigma\gamma_\iota\Big)k_\beta p_\iota
\nonumber\\&&
+(k\leftrightarrow p)(\rho\leftrightarrow\sigma )
\nonumber\\&&
=  \frac{2}{{D}}
\int \frac{d^D q}{(2\pi)^D}
\frac{q^2
}{ (q^2-\Delta)^3}(D-4)
Tr\Big(\gamma_\chi \gamma_\rho \gamma_\beta 
\gamma_\sigma\gamma_\iota\Big)k_\beta p_\iota,
\label{anona.2.13}
\end{eqnarray}
where the dependence of $\Delta$ from $x,y$ has been
neglected due to the vanishing factor $D-4$.
%
\subsection{The Triangle Anomaly}
The expression in eq. (\ref{anona.2.13}) provides the
anomaly in presence of two external vector
mesons. Only the pole part of the integral provides
a non vanishing result
\begin{eqnarray}&&
i~
(p+k)_\mu ( T_{\mu\rho\sigma}(k,p)+T_{\mu\sigma\rho}(p,k))
\nonumber\\&&
=\frac{2}{{D}} (-\frac{i}{(4\pi)^2})\frac{2}{D-4}(D-4)
Tr\Big(\gamma_\chi \gamma_\rho \gamma_\beta 
\gamma_\sigma\gamma_\iota\Big)k_\beta p_\iota
\nonumber\\&&
= {-} (\frac{i}{(4\pi)^2})
Tr\Big(\gamma_\chi \gamma_\rho \gamma_\beta 
\gamma_\sigma\gamma_\iota\Big)k_\beta p_\iota.
\label{anona.2.14}
\end{eqnarray}
Finally we add the group factor from eq. (\ref{anona.4})
\begin{eqnarray}&&
i~
(p+k)_\mu ( T^{ab}_{\mu\rho\sigma}(k,p)+T^{ab}_{\mu\sigma\rho}(p,k))
\nonumber\\&&
={-\frac{1}{2}} \delta_{ab} (\frac{i}{(4\pi)^2})
Tr\Big(\gamma_\chi \gamma_\rho \gamma_\beta 
\gamma_\sigma\gamma_\iota\Big)k_\beta p_\iota.
\label{anona.2.15}
\end{eqnarray}
In terms of fields this is
\begin{eqnarray}&&
\partial_\mu J_\mu^5
= {-\frac{1}{4}}(\frac{i}{(4\pi)^2})
Tr\Big(\gamma_\chi \gamma_\rho \gamma_\beta 
\gamma_\sigma\gamma_\iota\Big)
\partial_\beta A^a_\rho\partial_\iota A^a_\sigma 
\nonumber\\&&
={-} 
{\frac{1}{2}}(\frac{i}{(4\pi)^2})Tr\Big(\gamma_\chi \gamma_\rho \gamma_\beta 
\gamma_\sigma\gamma_\iota\Big)
tr\Big(
\partial_\beta A_\rho\partial_\iota A_\sigma \Big)
\label{anona.2.16}
\end{eqnarray}
which is in agreement with the result in I.

\section{One-loop Box Contribution}
\label{sec:box}
The amplitude for the box diagram {(by neglecting the
group factors)} is given by the Feynman rules $- i^4 \frac{i^3}{2^3}$
(four propagators, three vertices and a $-$ due to the fermion loop)
\begin{eqnarray}&&
T_{\mu\rho\sigma\nu}^{\rm Box}(k,p,l)=\frac{{i}}{2^3}
\int \frac{d^D q}{(2\pi)^D}~
 Tr ~\Bigl(\gamma_\mu{\gamma_\chi}
q_\alpha\gamma_\alpha  \gamma_\rho(1+\gamma_\chi) (q+k)_\beta\gamma_\beta
\nonumber\\&& 
\gamma_\sigma(1+\gamma_\chi)
(q+k+p)_\iota\gamma_\iota\gamma_\nu(1+\gamma_\chi) (q+k+p+l)_\delta\gamma_\delta 
\Bigr)
\nonumber\\&&
[
 q^2(q+k)^2 (q+k+p)^2(q+k+p+l)^2
]^{-1}
\label{anona.15.0}
\end{eqnarray}
where incoming momenta and polarizations are $(k,\rho),
~(p,\sigma)$ and $(l,\nu)$.
\par
\subsection{One-loop Box Contribution: the Divergence of the Current}
We include also the group factor $tr(t_at_bt_c)=\frac{i}{4}\varepsilon_{abc}$.
Thus the divergence of the current at one loop is
\begin{eqnarray}&&
i(p+k+l)_\mu T_{\mu\rho\sigma\nu}^{\rm BoxDiv}(k,p,l)\frac{i}{4}\varepsilon_{abc}
\nonumber\\&& 
=-i\frac{{\varepsilon_{abc}}}{2^5}
\int \frac{d^D q}{(2\pi)^D}~(p+k+l)_\mu 
 Tr ~\Bigl(\gamma_\mu~{\gamma_\chi}
q_\alpha\gamma_\alpha  ~
\nonumber\\&& 
\gamma_\rho(1+\gamma_\chi) ~(q+k)_\beta\gamma_\beta~
\gamma_\sigma~(1+\gamma_\chi)
(q+k+p)_\iota\gamma_\iota\gamma_\nu(1+\gamma_\chi) 
\nonumber\\&&
(q+k+p+l)_\delta\gamma_\delta 
\Bigr)
[
 q^2(q+k)^2 (q+k+p)^2(q+k+p+l)^2
]^{-1}
\nonumber\\&& 
=-i\frac{{\varepsilon_{abc}}}{2^5}
\int \frac{d^D q}{(2\pi)^D}   \Bigg\{
\nonumber\\&& 
- Tr ~\Bigl(~{q_\mu \gamma_\mu\gamma_\chi}~q_\alpha\gamma_\alpha  
\gamma_\rho(1+\gamma_\chi) ~(q+k)_\beta\gamma_\beta~
\gamma_\sigma~(1+\gamma_\chi)
(q+k+p)_\iota\gamma_\iota\gamma_\nu
\nonumber\\&& 
(1+\gamma_\chi) (q+k+p+l)_\delta\gamma_\delta
\Bigr)
[q^2
 (q+k)^2 (q+k+p)^2(q+k+p+l)^2
]^{-1}
\nonumber\\&&  
+Tr ~\Bigl(
 ~{\gamma_\chi} q_\alpha\gamma_\alpha 
\gamma_\rho(1+\gamma_\chi) ~(q+k)_\beta\gamma_\beta~
\gamma_\sigma~(1+\gamma_\chi)
\nonumber\\&& 
(q+k+p)_\iota\gamma_\iota\gamma_\nu(1+\gamma_\chi) 
\Bigr)
[
 q^2(q+k)^2 (q+k+p)^2
]^{-1}\Bigg\}
\label{anona.15.1}
\end{eqnarray}
The sum over the permutations of $(a,\rho, k)$, 
$(b,\sigma, p)$ and $(c, \nu, l)$ is understood.
\subsection{Identities at $D=4$}
It is convenient to disclose the identities
that would be satisfied in a situation where
$D$ can be taken equal to $4$. Thus we consider  
the first integral of the RHS where we identify the part
responsible for the anomaly (i.e. $\{\gamma_\mu,\gamma_\chi\}$)
of eq. (\ref{anona.15.1})
\begin{eqnarray}&&
-i\frac{{\varepsilon_{abc}}}{2^5}
\int \frac{d^D q}{(2\pi)^D}  
\nonumber\\&& 
- Tr ~\bigg(~\Big({\{\gamma_\mu,\gamma_\chi\}-\gamma_\chi\gamma_\mu}\Big)
q_\mu q_\alpha\gamma_\alpha  
\gamma_\rho(1+\gamma_\chi)
 ~ (q+k)_\beta \gamma_\beta~
\nonumber\\&& 
\gamma_\sigma~(1+\gamma_\chi)
(q+k+p)_\iota\gamma_\iota\gamma_\nu
(1+\gamma_\chi) (q+k+p+l)_\delta\gamma_\delta
\bigg)
\nonumber\\&& 
[q^2
 (q+k)^2 (q+k+p)^2(q+k+p+l)^2
]^{-1}
\label{anona.15.1.0}
\end{eqnarray}
The non anomalous part ($-\gamma_\chi\gamma_\mu$) should 
contribute to the cancellations in the divergence of the isoscalar
axial current. We elaborate this quantity by replacing
$q\to q -k$
\begin{eqnarray}&&
-i\frac{{\varepsilon_{abc}}}{2^5}
\int \frac{d^D q}{(2\pi)^D}  
\nonumber\\&& 
 Tr ~\bigg(\gamma_\chi
\gamma_\rho(1+\gamma_\chi)q_\beta \gamma_\beta~
\gamma_\sigma~(1+\gamma_\chi)
\nonumber\\&& 
(q+p)_\iota\gamma_\iota\gamma_\nu
(1+\gamma_\chi)(q+p+l)_\delta \gamma_\delta
\bigg)
[ (q+p)^2 (q+p+l)^2q^2]^{-1}
\label{anona.15.1.1}
\end{eqnarray}
We add the expression in eq. (\ref{anona.15.1.1}) to the second
term in eq. (\ref{anona.15.1})
on which  we perform the cyclic permutation 
$(a,\rho,k)\to (b,\sigma,p)\to (c,\nu,l)\to (a,\rho,k) $.
The result of this sum is 
\begin{eqnarray}&&
-i\frac{{\varepsilon_{abc}}}{2^5}
\int \frac{d^D q}{(2\pi)^D}  \Bigg\{
 Tr ~\bigg(\gamma_\chi
\gamma_\rho(1+\gamma_\chi)q_\beta \gamma_\beta~
\gamma_\sigma~(1+\gamma_\chi)
\nonumber\\&& 
(q+p)_\iota\gamma_\iota\gamma_\nu
(1+\gamma_\chi)(q+p+l)_\delta \gamma_\delta
\bigg)
\nonumber\\&&  
+Tr ~\bigg(\gamma_\chi  q_\beta \gamma_\beta~
\gamma_\sigma(1+\gamma_\chi)(q+p)_\iota\gamma_\iota
\gamma_\nu~(1+\gamma_\chi)
\nonumber\\&& 
(q+p+l)_\delta\gamma_\delta\gamma_\rho
(1+\gamma_\chi)
\bigg)
\Bigg\}
[ (q+p)^2 (q+p+l)^2q^2]^{-1}
\nonumber\\&&
=-i\frac{{\varepsilon_{abc}}}{2^5}
\int \frac{d^D q}{(2\pi)^D} 
 Tr ~\bigg(\{\gamma_\chi,
\gamma_\rho\}(1+\gamma_\chi)q_\beta \gamma_\beta~
\gamma_\sigma~(1+\gamma_\chi)
\nonumber\\&& 
(q+p)_\iota\gamma_\iota\gamma_\nu
(1+\gamma_\chi)(q+p+l)_\delta \gamma_\delta
\bigg)
[ (q+p)^2 (q+p+l)^2q^2]^{-1}.
\label{anona.15.1.2}
\end{eqnarray}
We see that the expression in eq. (\ref{anona.15.1.2})
is vanishing if $\{\gamma_\chi,\gamma_\rho\}=0$.
\par
The same result can be obtained for all terms generated
from eq. (\ref{anona.15.1}) by using the permutations on the
external variables  
$(a,\rho,k)$, $ (b,\sigma,p)$ and $ (c,\nu,l)$.
\subsection{The Box Anomaly}
From the previous calculation we get the final result
for the anomaly coming from the box.
It is given by the sum over all permutations on
the external vector mesons of the term proportional
to $\{\gamma_\mu,\gamma_\chi\}$ in eq. (\ref{anona.15.1.0}) and 
of the expression in eq. (\ref{anona.15.1.2})
\begin{eqnarray}&&
-i\frac{{\varepsilon_{abc}}}{2^5}
\int \frac{d^D q}{(2\pi)^D}  \Bigg\{
- Tr ~\bigg({\{\gamma_\mu,\gamma_\chi\}}
q_\mu q_\alpha\gamma_\alpha  
\gamma_\rho(1+\gamma_\chi)
 ~ (q+k)_\beta \gamma_\beta~
\nonumber\\&& 
\gamma_\sigma~(1+\gamma_\chi)
(q+k+p)_\iota\gamma_\iota\gamma_\nu
(1+\gamma_\chi) (q+k+p+l)_\delta\gamma_\delta
\bigg)
\nonumber\\&& 
[q^2
 (q+k)^2 (q+k+p)^2(q+k+p+l)^2
]^{-1}
\nonumber\\&& 
+
 Tr ~\bigg(\{\gamma_\chi,
\gamma_\rho\}(1+\gamma_\chi)q_\beta \gamma_\beta~
\gamma_\sigma~(1+\gamma_\chi)
\nonumber\\&& 
(q+p)_\iota\gamma_\iota\gamma_\nu
(1+\gamma_\chi)(q+p+l)_\delta \gamma_\delta
\bigg)
[ (q+p)^2 (q+p+l)^2q^2]^{-1}
\Bigg\}.
\label{anona.15.1.3}
\end{eqnarray}
%
%
\subsection{The First Term in Eq. (\ref{anona.15.1.3})}
Let us consider the first term in eq. (\ref{anona.15.1.3}).
Since $\{\gamma_\mu,\gamma_\chi\}= {\cal O}(D-4)$ the gamma trace
reduces to 
\begin{eqnarray}&&
 Tr ~\bigg({\{\gamma_\mu,\gamma_\chi\}}
\gamma_\alpha  
\gamma_\rho(1+\gamma_\chi)
 \gamma_\beta~
\gamma_\sigma~(1+\gamma_\chi)
\gamma_\iota\gamma_\nu
(1+\gamma_\chi)\gamma_\delta
\bigg)
\nonumber\\&& 
=
4 Tr ~\bigg({\{\gamma_\mu,\gamma_\chi\}}
\gamma_\alpha  \gamma_\rho \gamma_\beta~\gamma_\sigma~\gamma_\iota\gamma_\nu
\gamma_\delta(1+\gamma_\chi)
\bigg).
\label{anona.15.1.4}
\end{eqnarray}
Let us focus now on the momentum integration.
Only the divergent
part of the $q-$integral can yield a non-zero result; i.e. the 4-th powers
of $q$ in the numerator. After Feynman parameterization, shift by
$q_\mu \to q_\mu + r_\mu$ and symmetric integration we get
\begin{eqnarray}
q_\mu q_\alpha q_\beta q_\iota
\to
\frac{q^4}{D(D+2)}\Big [\delta_{\mu\alpha}\delta_{\beta\iota}
+\delta_{\mu\iota}\delta_{\beta\alpha}
+\delta_{\mu\beta}\delta_{\iota\alpha}
\Big].
\label{anona.15.1.5}
\end{eqnarray}
Thus we can neglect the second $\gamma_\chi$ at the far right in eq. 
(\ref{anona.15.1.4}) and the numerator of the first term in eq. 
(\ref{anona.15.1.3}) after symmetric integration becomes
\begin{eqnarray}&&
- Tr ~\bigg({\{\gamma_\mu,\gamma_\chi\}}
q_\mu q_\alpha\gamma_\alpha  
\gamma_\rho(1+\gamma_\chi)
 ~ (q+k)_\beta \gamma_\beta~
\nonumber\\&& 
\gamma_\sigma~(1+\gamma_\chi)
(q+k+p)_\iota\gamma_\iota\gamma_\nu
(1+\gamma_\chi) (q+k+p+l)_\delta\gamma_\delta
\bigg)
\nonumber\\&& 
= - \frac{q^4}{D(D+2)}
4 Tr ~\bigg({\gamma_\chi\{\gamma_\mu,}
\gamma_\alpha  \gamma_\rho \gamma_\beta~\gamma_\sigma~\gamma_\iota\gamma_\nu
\gamma_\delta\}
\bigg)
\nonumber\\&& 
\Bigg(\Big [\delta_{\mu\alpha}\delta_{\beta\iota}
+\delta_{\mu\iota}\delta_{\beta\alpha}
+\delta_{\mu\beta}\delta_{\iota\alpha}
\Big](r+k+p+l)_\delta
\nonumber\\&& 
+\Big [\delta_{\mu\alpha}\delta_{\beta\delta}
+\delta_{\mu\delta}\delta_{\beta\alpha}
+\delta_{\mu\beta}\delta_{\delta\alpha}
\Big](r+k+p)_\iota
\nonumber\\&& 
+\Big [\delta_{\mu\alpha}\delta_{\iota\delta}
+\delta_{\mu\delta}\delta_{\iota\alpha}
+\delta_{\mu\iota}\delta_{\delta\alpha}
\Big](r+k)_\beta
\nonumber\\&& 
+\Big [\delta_{\mu\beta}\delta_{\iota\delta}
+\delta_{\mu\delta}\delta_{\iota\beta}
+\delta_{\mu\iota}\delta_{\delta\beta}
\Big] r_\alpha
\nonumber\\&& 
+\Big [\delta_{\alpha\beta}\delta_{\iota\delta}
+\delta_{\alpha\delta}\delta_{\iota\beta}
+\delta_{\alpha\iota}\delta_{\delta\beta}
\Big]r_\mu
\Bigg)
\label{anona.15.1.6}
\end{eqnarray}
We neglect the last line of eq. (\ref{anona.15.1.6}) since,
after the use of the Kronecker delta, too few gamma's are left
for a non-zero limit of $D=4$. Thus we have
\begin{eqnarray}&&
- Tr ~\bigg({\{\gamma_\mu,\gamma_\chi\}}
q_\mu q_\alpha\gamma_\alpha  
\gamma_\rho(1+\gamma_\chi)
 ~ (q+k)_\beta \gamma_\beta~
\nonumber\\&& 
\gamma_\sigma~(1+\gamma_\chi)
(q+k+p)_\iota\gamma_\iota\gamma_\nu
(1+\gamma_\chi) (q+k+p+l)_\delta\gamma_\delta
\bigg)
\nonumber\\&& 
= - \frac{q^4}{D(D+2)}
4 Tr ~\bigg({\gamma_\chi\Big\{\gamma_\mu,}
\nonumber\\&& 
\Big [
(2-D)\gamma_\mu  \gamma_\rho ~\gamma_\sigma~\gamma_\nu
\gamma_\iota
+(2-D) \gamma_\rho ~\gamma_\sigma~\gamma_\mu\gamma_\nu
\gamma_\iota
\nonumber\\&&
+(6-D)  \gamma_\rho \gamma_\mu~\gamma_\sigma~\gamma_\nu
\gamma_\iota
\Big](r+k+p+l)_\iota
\nonumber\\&& 
+\Big [(6-D)
\gamma_\mu  \gamma_\rho\gamma_\sigma~\gamma_\iota\gamma_\nu
+(2-D)  \gamma_\rho ~\gamma_\sigma~\gamma_\iota\gamma_\nu
\gamma_\mu
\nonumber\\&& 
+(10-D)
\gamma_\rho \gamma_\mu~\gamma_\sigma~\gamma_\iota\gamma_\nu
\Big](r+k+p)_\iota
\nonumber\\&&
+\Big [ (2-D)
\gamma_\mu  \gamma_\rho \gamma_\iota~\gamma_\sigma~\gamma_\nu
+(6-D)
 \gamma_\rho \gamma_\iota~\gamma_\sigma\gamma_\nu
\gamma_\mu
\nonumber\\&& 
+(10-D)
 \gamma_\rho \gamma_\iota~\gamma_\sigma~\gamma_\mu\gamma_\nu
\Big](r+k)_\iota
\nonumber\\&& 
+\Big [(2-D)
\gamma_\iota  \gamma_\rho \gamma_\mu~\gamma_\sigma~\gamma_\nu
+(2-D)
\gamma_\iota  \gamma_\rho ~\gamma_\sigma~\gamma_\nu
\gamma_\mu
\nonumber\\&& +(6-D)
\gamma_\iota  \gamma_\rho \gamma_\sigma~\gamma_\mu\gamma_\nu
\Big] r_\iota
\Big\}
\Bigg)
%
\nonumber\\&& 
= -{2(D-4)} \frac{q^4}{D(D+2)}
4 Tr ~\big({\gamma_\chi \gamma_\rho ~\gamma_\sigma~\gamma_\nu
\gamma_\iota\big )}
\nonumber\\&& 
\Bigg(
\Big [
(2-D) 
+(2-D) 
-(6-D)  
\Big](r+k+p+l)_\iota
\nonumber\\&& 
+\Big [-(6-D)
-(2-D) 
+(10-D)
\Big](r+k+p)_\iota
\nonumber\\&&
+\Big [ (2-D) +(6-D)
-(10-D)
\Big](r+k)_\iota
\nonumber\\&& 
-\Big [(2-D) +(2-D) -(6-D)
\Big] r_\iota
\Bigg)
\nonumber\\&& 
=-{2(D-4)} (D+2)\frac{q^4}{D(D+2)}
4 Tr ~\big({\gamma_\chi \gamma_\rho ~\gamma_\sigma~\gamma_\nu
\gamma_\iota\big )}
\nonumber\\&& 
\Big(-(r+k+p+l)
+(r+k+p)
-(r+k)_\iota + r
\Big)_\iota
\nonumber\\&& 
={8(D-4)}\frac{q^4}{D}
 Tr ~\big({\gamma_\chi \gamma_\rho ~\gamma_\sigma~\gamma_\nu
\gamma_\iota\big )}(k+l)_\iota
\label{anona.15.1.7}
\end{eqnarray}
%
\subsection{The Second Term in Eq. (\ref{anona.15.1.3})}
The second term in eq. (\ref{anona.15.1.3}) has also
to be evaluated in the process of symmetric integration over
$q$ after the shift 
\begin{eqnarray}
q_\mu \to q_\mu + r_\mu.
\label{anona.15.1.8}
\end{eqnarray}
Thus we have
\begin{eqnarray}
q_\mu  q_\nu \to \frac{q^2}{D}\delta_{\mu\nu}.
\label{anona.15.1.9-1}
\end{eqnarray}
We have 
\begin{eqnarray}&&
 Tr ~\bigg(\{\gamma_\chi,
\gamma_\rho\}(1+\gamma_\chi)q_\beta \gamma_\beta~
\gamma_\sigma~(1+\gamma_\chi)
\nonumber\\&& 
(q+p)_\iota\gamma_\iota\gamma_\nu
(1+\gamma_\chi)(q+p+l)_\delta \gamma_\delta
\bigg)
\nonumber\\&& 
= 4 
 Tr ~\bigg(\{\gamma_\chi,
\gamma_\rho\} \gamma_\beta~
\gamma_\sigma~
\gamma_\iota\gamma_\nu
(1+\gamma_\chi)\gamma_\delta
\bigg)
\nonumber\\&& 
\Big[ (q+r)_\beta(q+r+p)_\iota(q+r+p+l)_\delta 
\Big]
\label{anona.15.1.9}
\end{eqnarray}
After symmetric integration the second $\gamma_\chi$
in the RHS of eq. (\ref{anona.15.1.9}) can be neglected
by following the argument in Section \ref{sec:alg}
\begin{eqnarray}&&
 Tr ~\bigg(\{\gamma_\chi,
\gamma_\rho\}(1+\gamma_\chi)q_\beta \gamma_\beta~
\gamma_\sigma~(1+\gamma_\chi)
\nonumber\\&& 
(q+p)_\iota\gamma_\iota\gamma_\nu
(1+\gamma_\chi)(q+p+l)_\delta \gamma_\delta
\bigg)
\nonumber\\&& 
= 4 \frac{q^2}{D}
 Tr ~\bigg(\gamma_\chi\{
\gamma_\rho, \gamma_\beta~
\gamma_\sigma~
\gamma_\iota\gamma_\nu\gamma_\delta\}
\bigg)
\nonumber\\&& 
\Big[ \delta_{\beta\iota}(r+p+l)_\delta 
+\delta_{\iota\delta}r_\beta
+\delta_{\beta\delta}(r+p)_\iota
\Big]
\label{anona.15.1.10}
\end{eqnarray}
We evaluate the Kronecker delta's
\begin{eqnarray}&&
 Tr ~\bigg(\{\gamma_\chi,
\gamma_\rho\}(1+\gamma_\chi)q_\beta \gamma_\beta~
\gamma_\sigma~(1+\gamma_\chi)
\nonumber\\&& 
(q+p)_\iota\gamma_\iota\gamma_\nu
(1+\gamma_\chi)(q+p+l)_\delta \gamma_\delta
\bigg)
\nonumber\\&& 
= 4  \frac{q^2}{D}
 Tr ~\bigg(\gamma_\chi\bigg\{
\gamma_\rho, 
\Big[ (2-D)
\gamma_\sigma~\gamma_\nu\gamma_\delta(r+p+l)_\delta 
\nonumber\\&& 
+(2-D)\gamma_\beta~
\gamma_\sigma~
\gamma_\nu r_\beta
+ (6-D)
\gamma_\sigma~
\gamma_\iota\gamma_\nu(r+p)_\iota
\Big]\bigg\}
\bigg)=0
\label{anona.15.1.11}
\end{eqnarray}
around $D=4$.
\section{Anomaly from the Box}
By restoring the initial factor of eq. (\ref{anona.15.1.3})
the anomaly in the current conservation is
\begin{eqnarray}&&
-i\frac{{\varepsilon_{abc}}}{2^5}
\int \frac{d^D q}{(2\pi)^D}  \Bigg\{
{8(D-4)}\frac{q^4}{D}
 Tr ~\big({\gamma_\chi \gamma_\rho ~\gamma_\sigma~\gamma_\nu
\gamma_\iota\big )}(k+l)_\iota
\Bigg\}(q^2-\Delta)^{-4}
\nonumber\\&&
=-i
\frac{{\varepsilon_{abc}}}{2^5D}
(\frac{i}{(4\pi)^2}) \frac{2}{D-4}8(D-4)
 Tr ~\big({\gamma_\chi \gamma_\rho ~\gamma_\sigma~\gamma_\nu
\gamma_\iota\big )}(k+l)_\iota
\nonumber\\&&
=
\frac{{\varepsilon_{abc}}}{2D}
(\frac{1}{(4\pi)^2}) 
 Tr ~\big({\gamma_\chi \gamma_\rho ~\gamma_\sigma~\gamma_\nu
\gamma_\iota\big )}(k+l)_\iota
\nonumber\\&&
.
\label{anona.15.1.13}
\end{eqnarray}
The sum over the permutations at ($D=4$) gives
\begin{eqnarray}&&
{\varepsilon_{abc}}
(\frac{1}{(4\pi)^2}) 
\frac{2}{D}
 Tr ~\big({\gamma_\chi \gamma_\rho ~\gamma_\sigma~\gamma_\nu
\gamma_\iota\big )}(k+p+l)_\iota\, .
\label{anona.15.1.14}
\end{eqnarray}
In terms of fields we have
\begin{eqnarray}&&
\partial_\mu J^5_\mu
= 
(\frac{1}{(4\pi)^2}) 
\frac{1}{D}
 Tr ~\big({\gamma_\chi \gamma_\rho ~\gamma_\sigma~\gamma_\nu
\gamma_\iota\big )}(-4i) tr\Big(i\partial_\iota A_\rho A_\sigma A_\nu \Big)
\nonumber\\&&
= 
(\frac{1}{(4\pi)^2}) 
 Tr ~\big({\gamma_\chi \gamma_\rho ~\gamma_\sigma~\gamma_\nu
\gamma_\iota\big )} tr\Big(\partial_\iota A_\rho A_\sigma A_\nu \Big)
\label{anona.15.1.15}
\end{eqnarray}
where $tr$ is the trace over the $SU(2)$ internal indices. 
\par
Together eqs. (\ref{anona.2.16}) and  (\ref{anona.15.1.15})
give the anomaly in the covariant form
\begin{eqnarray}&&
\partial_\mu J^5_\mu
= 
(\frac{1}{(4\pi)^2}) \frac{i}{{8}}Tr ~\big({\gamma_\chi\gamma_\beta \gamma_\rho 
~\gamma_\iota\gamma_\sigma~\big )}tr\Big(G_{\beta\rho}G_{\iota\sigma}\Big)
\label{anona.15.1.16}
\end{eqnarray}
where 
\begin{eqnarray}
G_{\mu\nu} = \partial_\mu A_\mu - \partial_\nu A_\mu + i[A_\mu,A_\nu].
\label{anona.15.1.17}
\end{eqnarray}
%

\section{Conclusions}

The present analytic calculation of the anomaly
of the axial isoscalar current in the $SU(2)$
chiral theory indicates that a consistent definition
of the trace with
$\gamma_5$ in dimensional regularization is at hand.
\par
In this work we used the ingredients expected to be present in a
consistent solution of the problem: associative algebra
for the gamma's, Lorentz covariance, cyclicity,
smooth limit at $D=4$
\begin{eqnarray}
Tr(p)
= \sum _{h=0} A_h(p) (D-4)^h.
\label{intr.02p}
\end{eqnarray}
$Tr(p)$ is any trace of gamma's and $\gamma_\chi$, where
the Lorentz indices are all saturated by vectors and tensors
(e.g. $\delta_{\mu\nu}$) and $ A_h(p)$ are $D=4$ Lorentz invariants
(being $\varepsilon_{\alpha\beta\rho\sigma}$ allowed).
\par
The outlook is the extension of the integral representation
of the trace, discussed in a previous paper (I),  
to the situation where more than one  $\gamma_5$ is
present.

\section*{Acknowledgements}
%
We  gratefully acknowledge the warm hospitality of the 
Department of Physics of the University of Pisa
and of the INFN, Sezione di Pisa. We are thankful to 
Peter Breitenlohner and to Mario Raciti
for stimulating discussions.

\normalsize

\bibliography{reference}

\end{document}